\documentclass[useAMS,usenatbib]{mn2e}

\usepackage[dvipdfm]{graphicx,color}
\usepackage{amsmath,amssymb}
\usepackage{hyperref}

\def\ha{{\rm\,H$\alpha$}}
\def\hb{{\rm\,H$\beta$}}
\def\hg{{\rm\,H$\gamma$}}
\def\oii{{\rm\,[O{\sc ii}]}}
\def\sii{{\rm\,[S{\sc ii}]}}
\def\nii{{\rm\,[N{\sc ii}]}}
\def\oiii{{\rm\,[O{\sc iii}]}}
\def\hii{{\rm\,H{\sc ii}}}

\def\pa{{\rm\,Pa$\alpha$}}
\def\msun{{\rm M}$_{\odot}$}
\def\mn{\ifmmode {\mu {\rm m} ~}\else {$\mu $m ~}\fi}
\def\aj{AJ}%
\def\araa{ARA\&A}%
\def\apj{ApJ}%
\def\mnras{MNRAS}%
\def\pasp{PASP}%
\def\pasj{PASJ}%
\title[Correlation between star formation and electron density]
{Correlation between star formation activity and electron density of 
ionized gas at $\mathbf{z}$=2.5}
\author[R. Shimakawa et al.]
{Rhythm Shimakawa$^{1,2}$\thanks{E-mail: 
rhythm@naoj.org}, Tadayuki Kodama$^{2,3}$, 
Charles C. Steidel$^{4}$, Ken-ichi Tadaki$^{5}$, 
\newauthor Ichi Tanaka$^{1}$, Allison L. Strom$^{4}$, 
Masao Hayashi$^{3}$, Yusei Koyama$^{6}$, 
\newauthor Tomoko L. Suzuki$^{2}$, and Moegi Yamamoto$^{2}$
\\
$^{1}$Subaru Telescope, National Astronomical Observatory of Japan, N. A$\textquoteright$ohoku Pl., Hilo, HI 96720, USA \\
$^{2}$Department of Astronomical Science, SOKENDAI (The Graduate University for Advanced Studies), Osawa, Mitaka, Tokyo 181-8588, Japan \\
$^{3}$Optical and Infrared Astronomy Division, National Astronomical Observatory, Mitaka, Tokyo 181-8588, Japan \\
$^{4}$Department of Astronomy, California Institute of Technology, E. California Blvd., Pasadena, CA 91125, USA \\
$^{5}$Max-Planck-Institut f\"{u}r Extraterrestrische Physik, Giessenbachstrasse, D-85748 Garching, Germany \\
$^{6}$Institute of Space Astronomical Science, Japan Aerospace Exploration Agency, Sagamihara, Kanagawa, 252-5210, Japan \\
}
\begin{document}

\date{Accepted 2015 April 22.  Received 2015 March 25; in original form 2014 November 5}

\pagerange{\pageref{firstpage}--\pageref{lastpage}} \pubyear{2014}

\maketitle

\label{firstpage}

\begin{abstract}

In the redshift interval of $2<z<3$, the physical conditions 
of the inter-stellar medium (ISM) in star-forming galaxies 
are likely to be different from those in the local Universe 
because of lower gaseous metallicities, higher gas 
fractions, and higher star formation activities.
In fact, observations suggest that higher electron 
densities, higher ionization parameters, and harder UV 
radiation fields are common.

In this paper, based on the spectra of \ha-selected 
star-forming galaxies at $z=2.5$ taken with Multi-Object 
Spectrometer for InfraRed Exploration (MOSFIRE) on
Keck-1 telescope, 
we measure electron densities ($n_e$) using the oxygen line 
ratio (\oii$\lambda\lambda$3726,3729), and investigate the 
relationships between the electron density of ionized gas 
and other physical properties. 
As a result, we find that the specific star formation rate 
(sSFR) and the surface density of SFR 
($\Sigma_\mathrm{SFR}$) are correlated with the electron 
density at $z=2.5$ for the first time.
The $\Sigma_\mathrm{SFR}$--$n_e$ relation is likely 
to be linked to the star formation law in \hii\ regions 
(where star formation activity is regulated by interstellar 
pressure). 
Moreover, we discuss the mode of star formation in those galaxies. 
The correlation between sSFR and $\Sigma_\mathrm{SFR}$ suggests 
that highly star-forming galaxies (with high sSFR) tend to 
be characterized by higher surface densities of star formation
($\Sigma_\mathrm{SFR}$) and thus higher $n_e$ values as well.

\end{abstract}


\section{Introduction}
After the recent advent of high sensitivity, multi-object 
near-infrared (NIR) spectrographs typified by MOSFIRE on 
the Keck telescope and KMOS (The K-band Multi Object 
Spectrograph) on the Very Large Telescope, physical 
properties of high-$z$ star-forming (SF) galaxies explored 
by deep spectroscopy have been hot issues with the initial 
burst of available data that are being accumulated 
\citep{Holden:2014, Steidel:2014, Shapley:2015, Coil:2015}. 
Most work has focused on the mass--metallicity 
relations \citep{Tremonti:2004, Sanders:2015}, the improved 
BPT diagrams \citep{Baldwin:1981, Steidel:2014, Shapley:2015} 
and MEx diagrams \citep{Juneau:2011, Juneau:2014} of SF 
galaxies at high redshifts. 
These analyses are now producing good measurements of 
gaseous excitation levels and metallicities of those galaxies.
In particular, recent large on-going programs 
such as KBSS-MOSFIRE (Strom et al. in prep.) and MOSDEF 
\citep{Kriek:2014} will soon provide us with 
NIR spectra of $\sim$1000 galaxies, all with reasonably 
high-quality spectra. 

However, so far most published analyses have not 
investigated another important parameter, that is the 
electron density $n_e$ at $z>2$. 
The electron density can be estimated from 
the ratios of collisonally-excited metal lines 
\oii$\lambda\lambda3726,3729$ or 
\sii$\lambda\lambda6717,6731$ doublets \citep{Osterbrock:1974}. 
This is a reliable physical quantity that is free of issues 
with flux calibration, is largely independent of metallicity, 
and only weakly depends on electron temperature.
One matter for concern is that \sii\ line should come 
from outer part of \hii\ region or photo-dissociation 
region (PDR, \citealt{Hollenbach:1999}) compared to \oii\ 
line. 
Thus the \oii\ doublet is in favor of measuring $n_e$ 
accurately. 

Some recent measurements suggest that the electron densities 
of high-$z$ ($z$$>$2) galaxies tend to have 
$n_e\sim30-1000$ /cm$^3$ \citep{Lehnert:2009, Masters:2014, 
Steidel:2014}, a range in which $n_e$ is sensitive to \oii\ 
and \sii\ line ratios which then enables us to investigate 
electron densities in high-$z$ galaxies.
\citet{Shirazi:2014} have reported that high-$z$ galaxies 
tend to have higher $n_e$ than those of low-$z$ counterparts 
for given sSFR and stellar mass, based on the indirect 
measurement (see the details in the literature).

This paper presents $n_e$ measurements, based on 
the \oii\ doublet by deep NIR spectroscopy of SF galaxies 
at $z=2.5$ obtained with MOSFIRE.
The electron density provides information on ISM conditions 
on scales much smaller than most of the measurements of gas surface 
density used in the context of the Kennicutt-Schmidt (KS) 
relation \citep{Schmidt:1959, Kennicutt:1998,Kennicutt:2012}. 
In the past decade, \citet{Liu:2008} have suggested 
that ISM condition is controlled by star formation 
intensities, and \citet{Brinchmann:2008b} have shown the 
correlation between specific star formation rate (sSFR) 
and electron density from the 
\sii\ doublet lines at $z\sim0.1$ for the first time. 
However no statistical study of electron density of 
SF galaxies at $z>2$ has been conducted so far. 
We present here the first of such analysis and show some intriguing 
correlations found between star formation and electron density 
of ionized gas at $z>2$ with the reliable, direct 
\oii\ measurements.
We assume the cosmological parameters of $\Omega_M$=0.3, 
$\Omega_\Lambda$=0.7 and $H_0$=70 km/s/Mpc and employ 
a \citet{Chabrier:2003} stellar initial mass function (IMF). 


\section{Observation, analyses, and data}

\subsection{Observation and analyses}

We selected our MOSFIRE spectroscopic targets from
68 \ha\ emitters (HAEs) 
associated with a protocluster, USS 1558--003 at $z=2.53$ 
\citep{Hayashi:2012} (hereafter USS1558). 
These are constructed from our previous narrow-band (NB) 
imaging survey with NB2315 ($\lambda_\mathrm{center}$=2317 
nm and FWHM=26 nm) with MOIRCS (Multi-Object Infrared Camera 
and Spectrograph; \citealt{Ichikawa:2006, Suzuki:2008}) on 
the Subaru telescope on Mauna Kea, as a part of the 
MAHALO-Subaru project
(Mapping HAlpha and Lines of Oxygen with Subaru;
\citealt{Kodama:2013}). 
The HAEs are first selected as those showing excess in 
NB flux above the 3$\sigma$ level, and equivalent widths 
greater than 20\AA\ in the rest frame. 
We then apply a broad-band colour-colour diagram ($r'JK_s$) 
in order to reduce fore- or back-ground contamination 
from \oiii/\oii/\pa\ emitters at some other redshifts
(see more detail in \citealt{Hayashi:2012}). 
Moreover, we placed higher priorities on the 36 objects 
which had been confirmed spectroscopically by 
our previous NIR spectroscopic observations 
with MOIRCS \citep{Shimakawa:2014, Shimakawa:2015}.

Spectroscopic observations were carried out in June 2014 
with MOSFIRE, a near-infrared imager and spectrograph 
\citep{McLean:2010, McLean:2012} on the Keck-1 telescope 
on Mauna Kea. 
A mask was created with 26 slits (each 0"7 wide) 
including 20 main targets, 5 galaxies consisting of 
2 NB emitters deselected by colour-colour diagram in 
\citet{Hayashi:2012} or 3 objects at photometric 
redshifts of $z$=1--2 by {\sc easy} \citep{Brammer:2008}, 
and a star to monitor atmosphere conditions. 
The observation was set with the ABA'B' dither pattern 
of individual exposure time of 120s in both $J$- and 
$H$-bands. 
We integrated for 3hrs in $J$- and $H$-bands each 
under 0.5--0.7 arc sec seeing size (FWHM). 
The $J$- and $H$-band modes of MOSFIRE provide high 
spectral resolutions of R=3318 and 3660 in the range of 
$\lambda$=1.17--1.35 and 1.47--1.80 $\mu$m, respectively,  
enabling us to resolve the \oii\ doublet in $J$-band at 
$z$=2.5.

The spectra were reduced using the MOSFIRE Data Reduction 
Pipeline ({\sc mosfire-drp}), described in more detail by 
\citet{Steidel:2014}. 
This software uses standard data reduction procedures and
produces a scientific 2D spectrum for each target. 
The flux calibration was conducted by using a standard star,
HD 199217 with an A0V spectrum. 

\begin{figure*}
\centering
\includegraphics[width=140mm]{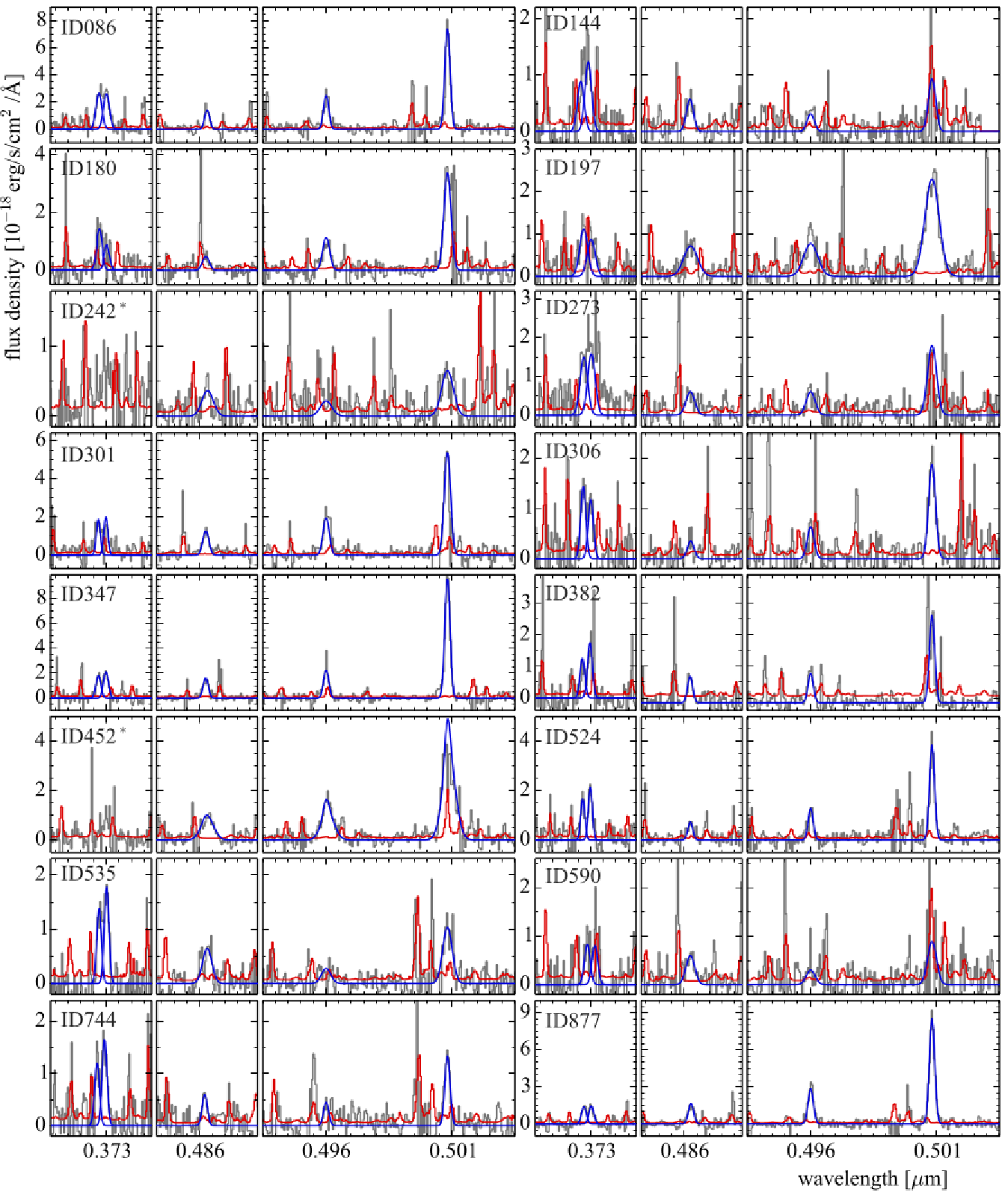}
\caption{Spectra of our HAEs at $z$$=$$2.5$.
In each panel, black, red and blue lines show the reduced 
spectrum, sky Poisson noise and the best fit curve to the
\oii$\lambda\lambda3726,3729$, \hb$\lambda4861$, and 
\oiii$\lambda\lambda4959,5007$ lines, respectively.
Two HAEs (ID242, 452) for which \oii\ doublet lines cannot be 
fit are excluded from our sample.}
\label{fig1}
\vspace{-15pt}
\end{figure*}

17 out of 20 of our target HAEs show more than one 
emission line, such as \oiii, \hb, \hg\ and \oii\ at the
protocluster redshift that are detected above the 5 sigma
flux limit. 
We have conducted a Gaussian fitting for the \oii\ and \oiii\ 
lines wherever available, using the spectral fitting software
{\sc specfit} \citep{Kriss:1994} distributed within 
{\sc stsdas}\footnotemark[1] in the {\sc iraf}\footnotemark[2]
environment. 
We assumed two Gaussian components with the same line width
to fit the respective \oii\ and \oiii\ lines at the fixed 
wavelengths of 3727.1/3729.9 \AA\ for the \oii\ doublet 
and 4960.3/5008.2 \AA\ for the \oiii\ doublet. 
We also fixed \oiii\ line ratio at 
\oiii$\lambda$5007/\oiii$\lambda$4959=3 \citep{Storey:2000}. 
The wavelength dependent sky Poisson noise that is largely
affected by the presence of OH sky lines is also taken into
account in the line fitting, and thus the fitting results 
should properly incorporate the observational uncertainties 
and the fitting errors. 
Obtained flux of these emission lines show the values 
between $0.5\times10^{-17}$ and $10\times10^{-17}$ 
erg/s/cm$^2$ and those S/N ratios are above 3.2$\sigma$. 
Typical line fluxes (and S/N levels) are 
\oiii$\lambda$5007=$3.6\times10^{-17}$ erg/s/cm$^2$ (15$\sigma$), 
\hb=$1.0\times10^{-17}$ erg/s/cm$^2$ (9$\sigma$) and 
\oii$\lambda\lambda$3726$+$3729=$2.6\times10^{-17}$ erg/s/cm$^2$ 
(11$\sigma$), respectively. 
Median 1$\sigma$ line flux uncertainties are 
1.6$\times$$10^{-18}$ and 1.8$\times$$10^{-18}$ erg/s/cm$^2$ 
in $J$ and $H$ bands. 
For 3 of 17 targets, one of the two \oii\ lines is severely 
affected by OH sky lines unfortunately, and thus the ratio 
could not be measured. 
Hereafter we do not include these galaxies in the analyses.
Our final sample therefore consists of 14 HAEs at $z=2.5$. 
All line spectra are presented in Fig. \ref{fig1}. 

\footnotetext[1]{Available at
www.stsci.edu/institute/software\_hardware/stsdas/}
\footnotetext[2]{{\sc iraf} is distributed by National 
Optical Astronomy Observatory and available at iraf.noao.edu/}

The electron density is then calculated for each HAE based 
on the measured intensity ratio of the \oii\ doublet using 
the {\sc temden} code distributed in the {\sc stsdas} package. 
The determination of the electron density depends weakly 
on the electron temperature, and we assume the fixed value 
of 10 000 K \citep{Osterbrock:1974} because independent 
measurements are not available. 
Note however that the choice of this value does
not affect our conclusions.

\subsection{Slit loss correction, stellar mass, and SFR}

\begin{figure}
\centering
\includegraphics[width=70mm]{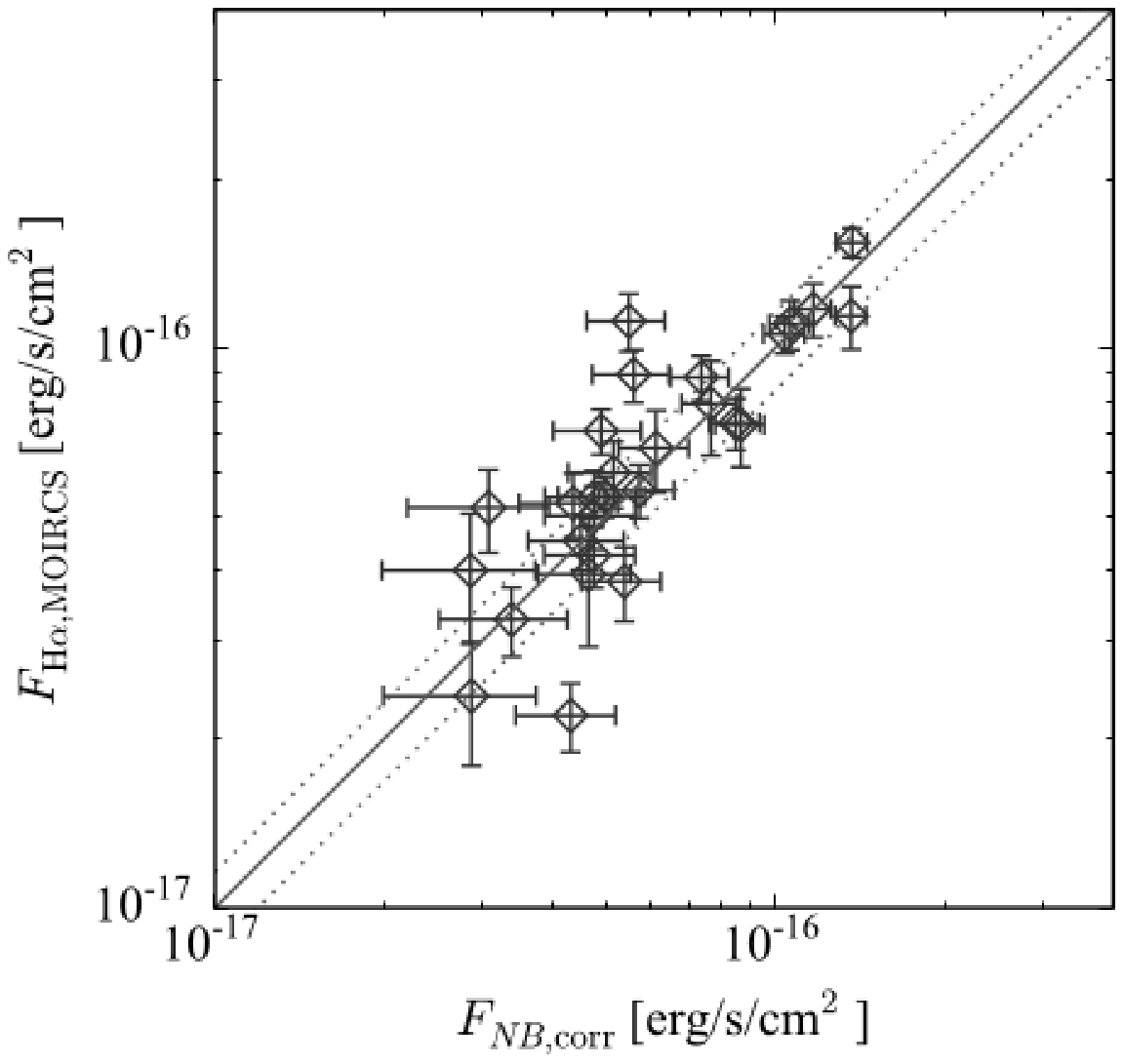}
\caption{Comparison between slit-loss corrected \ha\ line 
fluxes and \nii-corrected NB fluxes for HAEs at 
$z$=2.5 in \citet{Shimakawa:2015}. 
Solid and dotted lines show a 1:1 line and the median 
scatter of 16\%, respectively.
}
\label{fig2}
\end{figure}

In order to obtain absolute line flux $F_{int,obj}$, 
we correct for the slit loss by the following equation, 
\begin{equation}
F_{int,obj}=\frac{F_{int,star}}{F_{obs,star}}\times\frac{f_{loss,star}}{f_{loss,obj}}\times F_{obs,obj},
\end{equation}
where $F_{int,star}$ and $F_{obs,star}$ are the intrinsic flux 
derived from $J$- or $H$-band image and the observed flux of 
the monitoring star placed on our slit mask used for
spectroscopy. 
$f_{loss,star}$ is a slit loss of the monitoring star. 
Here we assume its value of 0.76 and 0.80 in $J$- and $H$-bands
which correspond to 0"70 and 0"65 of seeing size in FWHM,
respectively. 
$f_{loss,obj}$ is the slit loss of our science targets 
and is estimated by assuming the single Gaussian component
that is spilt outside of the 0"7 slit width for each FWHM
based on the NB image \citep{Hayashi:2012}.
$f_{loss,obj}$ is distributed between 0.42 and 0.78,
and a median value is 0.69. 
This correction may cause a large uncertainty for those 
having multiple \hii\ regions as pointed out 
by \citet{Kriek:2014}. 
Therefore we double-check this effect by comparing the
\nii-corrected NB fluxes with the spectroscopic
\ha\ line fluxes of 28 HAEs at $z=2.5$ (of which 10 are 
common in this work) measured from our previous MOIRCS 
spectra \citep{Shimakawa:2015}. 
The \nii\ correction is calibrated as a function of 
stellar mass as is applied in \citet{Shimakawa:2015}.
The result shows the typical uncertainty including the \nii\ 
correction is 16\% (Fig. \ref{fig2}). 
We ignore this small error hereafter, since this paper 
only discuss the statistical properties of the sample.

We use the photometric data ($Br'z'JHK_s$ and NB) catalog 
of \citet{Hayashi:2012} in order to obtain stellar masses 
and SFRs of the individual galaxies. 
The stellar masses are derived from the {\sc fast}, 
SED-fitting program \citep{Kriek:2009} using the stellar 
population model of \citet{Bruzual:2003}, the 
\citet{Calzetti:2000} extinction curve, and the 
\citet{Chabrier:2003} IMF. 
We fix the metallicity to the solar value for the SED models
and assume an delayed exponentially declining star formation 
history ($\sim$$t\exp(-t$/tau))
with the fixed time-scale of log($\tau$/yr)=8.4. 
The SFRs are calculated from \ha\ luminosity by using 
the \citet{Kennicutt:1998} prescription but with
\citet{Chabrier:2003} IMF. 
We have corrected for the nebular dust extinction based on the 
Balmer decrement (\ha/\hb=2.86) by assuming a Case-B 
recombination in the gas temperature of $T_e$=10$^4$ K 
and the electron density of $n_e$=10$^2$ cm$^{-3}$ 
\citep{Brocklehurst:1971}, as well as assuming the 
\citet{Calzetti:2000} extinction curve for \oii\ and 
\oiii\ lines. 
\ha\ and \hb\ fluxes are derived from the \nii-corrected 
NB flux and the spectroscopic \hb\ line flux whose slit 
loss is rectified as mentioned above. 


\section{Results}

\begin{figure}
\centering
\includegraphics[width=75mm]{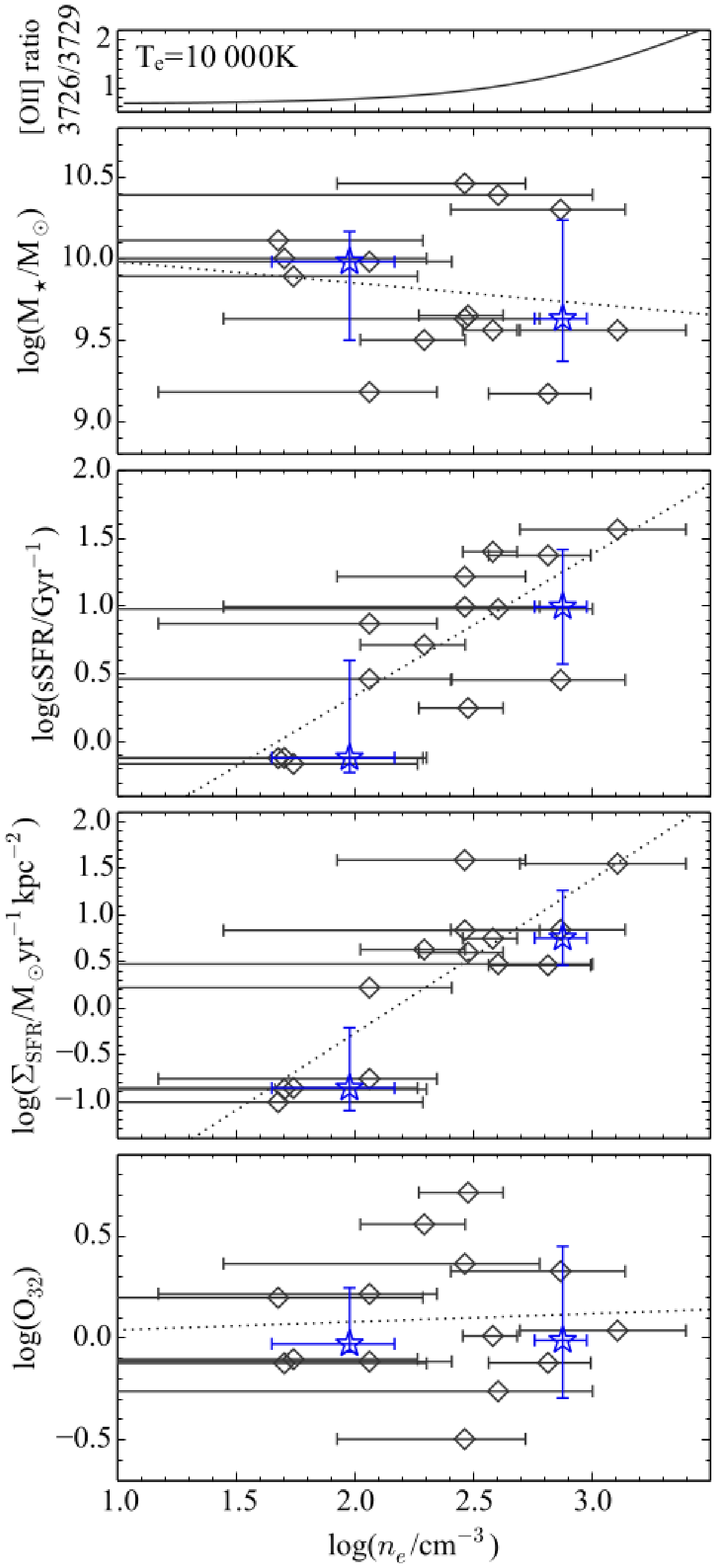}
\caption{From top, log (a) \oii$\lambda\lambda3726,3729$ 
line ratio, (b) stellar mass, (c) sSFR, (d) 
$\Sigma_\mathrm{SFR}$ and (e) O$_{32}$ as a function 
of log electron density. 
Diamonds indicate our sample, HAEs at $z=2.5$.
Errorbars of $n_e$ are determined from flux errors of 
\oii$\lambda\lambda3726,3729$.  
Blue star symbols represent $n_e$ of the stacked data 
separated by $n_e=160$ cm$^{-3}$ and those $y$-values 
show median values and 1$\sigma$ scatters. 
Dotted lines show the best fit lines for our targets. 
}
\label{fig3}
\end{figure}

Grounded on the datasets, this {\it Paper} explores the 
relationships of electron density ($n_e$) in the ionized
regions with stellar mass, sSFR, SFR surface density
($\Sigma_\mathrm{SFR}$) and \oiii/\oii\ line ratio. 
$\Sigma_\mathrm{SFR}$ is defined by SFR/$\pi 
r_\mathrm{eff}^2$ where $r_\mathrm{eff}$ is the effective 
radius of a NB image profile for each object.
We derive $r_\mathrm{eff}$ from the NB image 
(seeing 0"4 FWHM) by using the {\sc galfit version 3.0}
code \citep{Peng.CY:2010}. 
Here, we employed only the NB data with good seeing sizes 
among the data used in \citet{Hayashi:2012}. 
The seeing-limited data may cause a large uncertainty 
of $r_\mathrm{eff}$ for the galaxies which remain 
unresolved in some cases. 
However, our conclusion remains the same even if we 
exclude those small, unresolved galaxies. 
The median effective radius is $r_\mathrm{eff}$=2.1
kpc which corresponds to 0.26 arc sec.

Figure \ref{fig3} represents (a) 
\oii$\lambda\lambda3726,3729$ line ratio, log (b) stellar 
mass, (c) sSFR, (d) surface density of SFR and (e) 
\oiii/\oii\ line ratio 
(O$_{32}$$\equiv$\oiii$\lambda\lambda4959,5007$/\oii$\lambda\lambda3726,3729$) 
as a function of log electron density. 
All physical parameters are dust-corrected by using 
the Balmer decrement and the \citet{Calzetti:2000} 
extinction curve (\S2.2). 
First of all, as seen in the Fig. \ref{fig3}a, 
the \oii$\lambda\lambda3726,3729$ line ratio is most 
sensitive for $n_e$$>$100 cm$^{-3}$. 
On the other hand, there is a large uncertainty in
estimating electron density for the galaxies with low
$n_e$ since \oii\ line ratio is saturated below
$\sim$10 cm$^{-3}$. 
Measured \oii$\lambda3726$/\oii$\lambda3729$ ratios 
show values between 0.71--1.6, and the median electron 
density is 291 cm$^{-3}$. 
To investigate the correlations among the physical
quantities, we calculate Spearman's rank-correlation 
coefficients $r_s$ and those significance. 
These are summarized in Table \ref{tab1}. 
Also, we check them excluding low mass galaxies 
(log(M$_\star$/\msun)$<$10) as well, in 
order to minimize the stellar mass dependence. 

\begin{table}
\centering
\caption{Fitted slopes ($N$) and those standard errors 
to the HAEs on log stellar mass, 
sSFR, $\Sigma_\mathrm{SFR}$ or O$_{32}$ ($y$) plotted 
against log electron density ($x\equiv$log($n_e$/cm${^3}$)). 
3rd and 4th columns indicate Spearman's rank 
correlation coefficients and their significance in 
the respective diagram suggested for the HAEs at $z=2.5$. 
}
 \begin{tabular}{@{}lrrr@{}}
 \hline
  $y$ & $N$ & $r_s$ & $\sigma$ \\
\hline
log(M$_\star$/M$_\odot$) & --0.13$\pm$0.26 & --0.16 & 1.3 \\
log(sSFR/Gyr$^{-1}$) & 1.04$\pm$0.23 & 0.68 & 4.0 \\
log($\Sigma_\mathrm{SFR}$/M$_\odot$yr$^{-1}$kpc$^{-2}$) & 1.65$\pm$0.29 & 0.73 & 4.2 \\
log(O$_{32}$) & 0.04$\pm$0.21 & 0.07 & 0.6 \\
\hline
\end{tabular}
\label{tab1}
\end{table}

In our sample, sSFR (Fig. \ref{fig3}c) and SFR surface 
density (Fig. \ref{fig3}d) are both correlated with 
electron density with $\sim4\sigma$ significance 
(Table \ref{tab1}). 
Current data show sSFR$\propto$$n_e^{1.1\pm0.2}$ and 
$\Sigma_\mathrm{SFR}$$\propto$$n_e^{1.7\pm0.3}$. 
These power laws change depending on absorption 
corrections although the tight relationships remain constant. 
We also can see the positive correlations in two composite 
spectra which are separated by $n_e$=160 cm$^{-3}$. 
These are stacked with weight of noise 
($\equiv\Sigma_i^n$($fluxdensity$/$noise^2$)/$\Sigma_i^n$(1/$noise)^{2}$). 
On the other hand, any dependencies of stellar mass 
and O$_{32}$ on electron density (Fig. \ref{fig3}b,e) 
cannot be identified due to a lack of statistics. 
However, we have confirmed a moderate 
negative correlation between log stellar mass and 
O$_{32}$ with 3$\sigma$ significance, which agrees 
with \citet{Nakajima:2014, Steidel:2014}. 


\section{Discussion and Conclusions}

\subsection{Star formation law in ionized gas}
Past works have indicated the association of 
star formation with electron density \citep{Martin:1997, 
Liu:2008, Brinchmann:2008b, Shirazi:2014}.
This work clearly shows quantitatively the correlation between
sSFR and electron density
and the one between $\Sigma_\mathrm{SFR}$ and electron density at $z=2.5$
based on the reliable measurements.
These relationships indicate that the interstellar 
pressure and the SF activities are closely linked. 
In particular, the $n_e$--$\Sigma_\mathrm{SFR}$ 
correlation should mean that the interstellar pressure 
controls the intensity of SF activity ($\Sigma_\mathrm{SFR}$) in 
\hii\ regions.
The electron density probes the physical 3D density 
of ionized SF regions, and it must be related to other 
2D density measurements of molecular or neutral gases. 
Also a typical \hii\ region has a lifetime of 
a few Myrs (comparable to the lifetime of OB stars) 
which is by a factor of a few -- an order of magnitude shorter than 
that of cold gas clouds. This means that the electron 
density more selectively corresponds to the interstellar pressure
of young star-forming regions.
If the electron density is somehow related to the density of cold ISM,
which is likely the case since the ionized gas originates from molecular gas,
the relation between $\Sigma_\mathrm{SFR}$ and electron density that we see here
must be closely related to the the KS law
(see also \citealt{Shirazi:2014}). 

However, we need some careful considerations for the 
implications of this correlation. 
First of all, it is not necessarily the case that $n_e$--$\Sigma_\mathrm{SFR}$ 
correlation links to the KS law ($\Sigma_\mathrm{gas}$--$\Sigma_\mathrm{SFR}$). 
The measured electron density directly scales with
the gas density if the hydrogen is fully-ionized and only 
in the SF \hii\ regions. 
However, the ionized gas are not only confined in the SF 
regions containing young massive stars, but also seen 
around AGNs or in the outer diffuse gas (warm ionized 
medium, \citealt{Oey:1997, Martin:1997, Charlot:2001}), 
and therefore the electron density does not necessarily 
correspond to the gas density of the SF regions. 
Secondly, \oii\ line fluxes that we obtain are 
the surface-brightness-weighted and ensemble averaged 
for the entire galaxy or the part that falls into the 
slit of 0"7 width.
In particular, the electron density is measured on quite 
a different scale (50--100 pc) from that of surface gas 
density used for the KS law ($\sim$1 kpc). 
In fact, the slope of 
$\Sigma_\mathrm{SFR}$--$\Sigma_\mathrm{gas}$ depends 
on the observed scale \citep{Oey:2007, Prescott:2007, 
Bigiel:2008, Heiderman:2010} and \citet{Kruijssen:2014} 
have theoretically shown the scale-dependence of the 
KS relation. 
Such combined effects of the surface-brightness-weighted measurements
and its scale dependence make the physical understanding of the
$n_e$--$\Sigma_\mathrm{SFR}$ correlation difficult.

These problems will be solved by integral field 
spectroscopy assisted by adaptive optics on next generation 
telescopes such as Thirty Meter Telescope whose spatial 
resolution is as good as 100 pc at $z\sim2$.
Alternatively, local galaxies with high 
$\Sigma_\mathrm{SFR}$ will be able to provide us with 
crucial information even under natural seeing conditions
by comparing the spatially-resolved electron density map
with that of HCN, for example, which traces dense clumps with 
forming stars. 

\subsection{Relationship between sSFR and $\mathbf{\Sigma_{SFR}}$}

\begin{figure}
\centering
\includegraphics[width=70mm]{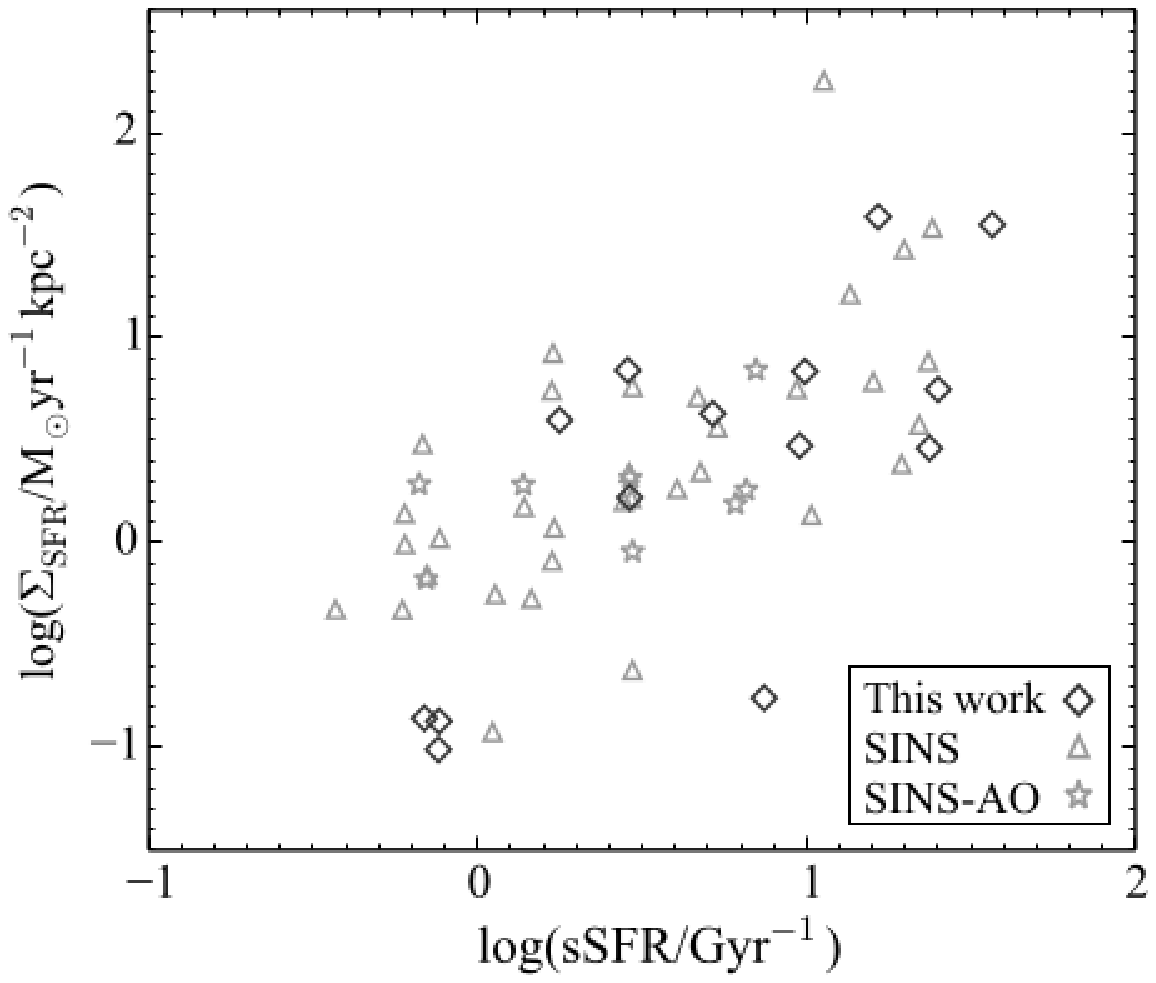}
\caption{Log sSFR versus SFR surface density. 
Along with the data of this work (the same plots as 
in Fig. \ref{fig3}), triangles and star marks represent
SF galaxies at $z\sim2$ in the SINS survey 
\citep{Schreiber:2009}. 
The star marks indicate those whose $r_\mathrm{eff}$
are derived from AO-assisted IFU observations. 
}
\label{fig4}
\end{figure}

Then, we discuss the mode of star formation in 
high-$z$ galaxies based on the sSFR, $\Sigma_\mathrm{SFR}$, 
and $r_\mathrm{eff}$, including the data from the SINS survey
(Spectroscopic Imaging survey in the near-IR with SINFONI) 
of high redshift galaxies \citep{Schreiber:2009}. 
This work employs their stellar masses and SFRs derived 
from the SED-fitting that are all available in the literature. 
We note that our data exactly follow the same 
main-sequence (i.e. relationships between stellar mass 
and SFR) of their sample. 
Therefore we analyse and plot their samples together with ours.

The results are shown in Figure \ref{fig4}. 
We can see a correlation between sSFR and SFR surface density
for all the samples. 
The Spearman's correlation factor is calculated to $r_s$=0.68
which shows above 5$\sigma$ significance. 
The sSFR indicates the strength of global SF activity like 
star-burstiness, while $\Sigma_\mathrm{SFR}$ represents
more local but averaged surface density of star forming regions.
Therefore this correlation suggests that the SF galaxies with 
higher sSFRs tend to be characterized by higher surface densities of
star formation rate ($\Sigma_\mathrm{SFR}$) probably due to
higher cold gas densities according to the KS law.
It is also probably connected to higher electron densities.
This may be understood if we consider that a burst-like intense
star formation tends to occur in a central compact region 
(see also \citealt{Wuyts:2011}).

\bigskip
Finally, we should note that our sample is biased to 
HAEs at $z=2.5$ in the protocluster environment. 
This means that environmental effects on feeding and 
feedback in galaxy formation may influence the chemical
evolution of SF galaxies even at the fixed stellar mass
\citep{Kulas:2013, Shimakawa:2015, Valentino:2015}.
It may cause a large scatter in O32, 
since O32 is not only dependent on ionization parameter, 
but also gaseous metallicity. 
Furthermore, it is expected that the mode of star formation 
depends on environment. 
For example, galaxy-galaxy interactions occur more 
frequently in cluster environment \citep{Gottlober:2001}. 
The model of SF in cluster galaxies could be different  
due to more frequent nuclear starbursts by major mergers. 
However, this effect is largely independent of the 
correlations found in this {\it Paper}, since our sample 
follow the same sequence of SF galaxies studied by SINS 
survey in the diagrams between sSFR and $\Sigma_\mathrm{SFR}$. 

\bigskip
In summary, this paper shows and discusses 
the star formation law in ionized gas characterized by 
the $n_e$--$\Sigma_\mathrm{SFR}$ correlation for the 
first time. 
So far, the accurate slope and its physical understanding 
remain unclear, and more statistical and coherent 
sample is required.
To achieve this, MOSFIRE is a powerful spectrograph 
which has good enough sensitivity to detect and resolve
\oii\ doublet lines for individual galaxies at high-$z$
at reasonably high S/N levels. 
Large programs such as KBSS, MOSDEF and the other on-going 
projects using MOSFIRE have already started and 
will be able to shed light in the physical states of star forming galaxies
at the peak epoch of galaxy formaiton.


\section*{Acknowledgments}
We are grateful to the anonymous referee for useful comments. 
The data presented in this paper were obtained at the 
W. M. Keck Observatory, which is operated as a scientific 
partnership among the California Institute of Technology, 
the University of California and the National Aeronautics and 
Space Administration. 
The Observatory was made possible by the generous 
financial support of the W.M. Keck Foundation.
Data analysis were in part carried out on common use data 
analysis computer system at the Astronomy Data Center, ADC, 
of the National Astronomical Observatory of Japan.


\bibliographystyle{mn2e}
\bibliography{bibtex_library}

\bsp

\label{lastpage}

\end{document}